\documentclass{elsart}
\usepackage[dvips]{graphicx}
\usepackage{epsfig,float}
\usepackage{graphics}

\begin{document}

\bibliographystyle{unsrt}


\newcommand  {\gtsim }
{\,\vcenter{\hbox{$\buildrel\textstyle>\over\sim$}}\,}
\newcommand  {\lesssim }
{\,\vcenter{\hbox{$\buildrel\textstyle<\over\sim$}}\,}
\newcommand  {\sech }  {\,\mathrm{sech} \,}
\newcommand  {\sign }  {\,\mathrm{sign} \,}

\newcommand{\parr}{\bigskip \bigskip  \bigskip}
\def\BE{\begin{equation}}
\def\EE{\end{equation}}
\def\BA{\begin{eqnarray}}
\def\EA{\end{eqnarray}}

\runauthor{Montagne,Hern{\'a}ndez--Garc{\'\i}a}
\begin{frontmatter}
\title{Localized structures in coupled Ginzburg--Landau equations}
\author[montagne]{Ra\'ul Montagne\cite{moemail}}
\author[emilio]{Emilio Hern{\'a}ndez-Garc{\'\i}a \cite{ememail}}

\address[montagne]{Instituto de F{\'\i}sica -- Universidad de la Rep{\'u}blica. \\
Igu{\'a} 4225, Montevideo 11400 --Uruguay.}\thanks{http://www.fisica.edu.uy/$\sim$montagne}
\address[emilio]{Instituto
Mediterraneo de Estudios Avanzados,
IMEDEA. \\
CSIC-Universitat de les Illes Balears, E-07071 Palma de Mallorca
--Spain.}\thanks{http://www.imedea.uib.es/Nonlinear}

\begin{abstract}
Coupled Complex Ginzburg-Landau equations describe generic
features of the dynamics of coupled fields when they are close to
a Hopf bifurcation leading to nonlinear oscillations. We study
numerically this set of equations and find, within a particular
range of parameters, the presence of uniformly propagating
localized objects behaving as coherent structures. Some of these
localized objects are interpreted in terms of exact analytical
solutions.
\end{abstract}
\end{frontmatter}

\section{Introduction}
\label{sec:Intro} When an extended system is close to a Hopf
bifurcation leading to uniform oscillations, the amplitude of the
oscillations can be generically described in terms of the complex
Ginzburg-Landau (CGL) equation \cite{CrossHohenberg}. When there
are two fields becoming unstable at the same bifurcation, coupled
complex Ginzburg-Landau equations (CCGL) should be used instead.
This model set of equations appears in a number of contexts
including convection in binary mixtures and transverse
instabilities in unpolarized lasers
\cite{CrossHohenberg,maxi95,gil93}.

Coherent structures such as fronts, shocks, pulses, and other
localized objects play an important role in the dynamics of
extended systems \cite{riecke98}. In particular, for the complex
Ginzburg-Landau equation, they provide the {\sl building blocks}
from which some kinds of spatiotemporally chaotic behavior are
built-up \cite{hecke98}. A systematic study of localized structures
in CCGL equations in one spatial dimension was initiated in
\cite{saarloos98}.

Here we present results on one dimensional CCGL equations in
parameter ranges such that
they can be written as
\begin{equation}
\partial_t A_{\pm}
= \mu A_{\pm} + (1 + i \alpha) \partial_x^2 A_{\pm}
 - (1+i\beta) \left( |A_{\pm}|^2 + \gamma |A_{\mp}|^2
\right) A_{\pm} \label{theEqu}
\>.
\end{equation}
Group velocity terms of the form $\pm v_g\partial_x A_\pm$ are
explicitly excluded, and $\gamma$ is restricted to take real
values (without additional loss of generality, $\alpha$ and
$\beta$ are also real parameters). In addition we just consider
$1+\alpha\beta>0$ (Benjamin-Feir stable range). These restrictions
are the appropriate ones for the description of transverse laser
instabilities \cite{maxi95}. In that case $A_\pm$ are related to
the two orthogonal circularly polarized light components. We
further restrict our study to the case $0<\gamma<1$ which is the
range obtained when atomic properties in the laser medium favor
linearly polarized emission. In terms of the wave amplitudes
$A_\pm$, wave coexistence is preferred.

\section{Numerical studies}

Many experiments on traveling wave systems or numerical
simulations of Ginzburg--Landau--type equations
\cite{CrossHohenberg,montagne96c,montagne98c} exhibit local
structures that have a shape essentially time--independent and
propagate with a constant velocity, at least during an interval of
time where they appear to be coherent structures
\cite{hohenbergsaarloos,montagne96c,saarloos98}. In order to
analyze these structures it is common to reduce the initial
partial differential equation to a set of ordinary differential
equations by restricting the class of solutions to uniformly
traveling ones. Localized structures are homoclinic or
heteroclinic orbits in this reduced dynamical system, that is they
approach simple solutions (typically plane waves) in opposite
parts of the system, whereas they exhibit a distinct shape in
between.

Instead of looking for solutions of the reduced dynamical system,
we prefer here to resort to direct numerical solution of
(\ref{theEqu}) under different initial conditions. A
pseudo--spectral code \cite{montagne96c,hoyuelos99} with periodic
boundary conditions and a second--order accuracy in time is used.
Spatial resolution was typically 512 modes. Time step was
typically $0.05$. The system size was always taken to be $L =
512$. Several kinds of localized objects which maintain coherence
for a time appear and travel around the system. Different initial
conditions give birth to different kinds of structures. Some of
them decay shortly, and the qualitative dynamics at long times
becomes determined by the remaining ones, and essentially
independent of the initial conditions.

The upper part of Fig. 1 shows the spatiotemporal
evolution of $|A_{+}(x,t)|$ and $|A_{-}(x,t)|$ at parameter values
$ \alpha = -0.35 \, ,\, \beta = -2.0 \mbox{ and } \gamma = 0.2$. Time
runs upwards and $x$ is represented in the horizontal direction.
Lighter grey corresponds to the maximum values of $|A_{\pm}(x,t)|$
and darker to the minima.
\begin{figure}[H]
\begin{center}
\resizebox{80mm}{!}{\includegraphics{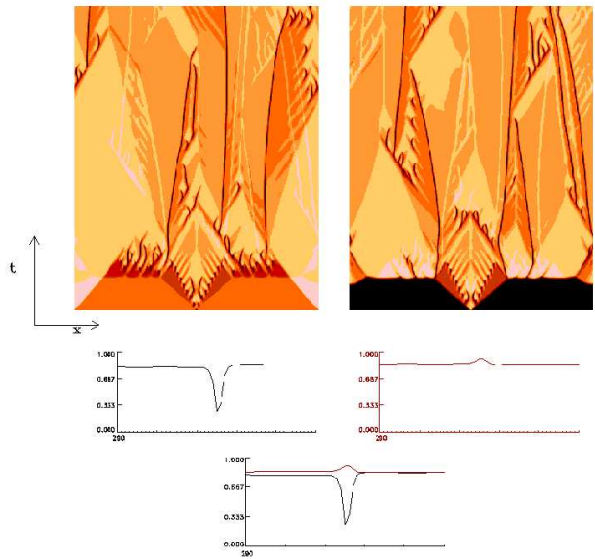}}
\end{center}
\caption[2]{Upper panels: Spatiotemporal evolution of $|A_{+}(x,t)|$
and $|A_{-}(x,t)|$ with time running upwards from $t=0$ to $400$
and $x$ in the horizontal direction, from $x=0$ to $x=512$.
Lighter grey corresponds to the maximum value of $|A_{\pm}(x,t)|$
and darker to  the minimum. Parameter values are $\alpha=-0.35$,
$\beta=-2.0$, and $\gamma=0.2$. Lower panels: A coupled
hole--maximum pair at $t= 399$ close to $x\approx 300$. This is
the dominant coherent structure at long times. Left
$|A_{+}(x,t)|$, right $|A_{-}(x,t)|$, and both graphs are
superposed in the central bottom panel.}
\label{fig297c}
\end{figure}

This particular evolution was obtained
starting from $A_{+}(x,0)$ equal to the \textsl{Nozaki-Bekki
hole}, a known analytical solution of the single Ginzburg-Landau
equation \cite{bekki83,bekki84}, and for $A_{-}(x,0)$ a
\textsl{Nozaki-Bekki pulse} \cite{bekki84}. These are not exact
solutions of the set of equations (\ref{theEqu}) so that this
initial condition decays and gives rise to complex spatiotemporal
structures. After a transient that will be described below, the
configuration of the system consists in portions with a modulus
nearly constant (corresponding to plane wave states) interrupted
by localized objects with particle-like behavior. Dark features in
$|A_+|$ appear where $|A_-|$ has bright features, thus indicating
that the localized object carries a kind of anticorrelation
between the fields. The lower panels of  Fig. 1 show the modulus
of the two fields at $t=399$ and $x \approx 300$, where one of
such objects is present. One of the components shows a maximum in
the modulus, whereas the other displays a deep minimum. We can
call this object a ``hole--maximum pair". It seems to be a
dissipative analog of the `out-gap' solitons appearing in Kerr
media with a grating \cite{feng93}, and here it is the
characteristic object building-up the disordered intermittent
dynamics seen at long times. It is clear that these objects
connect the plane wave states (that is the constant modulus
regions) filling most of the system.  Before reaching the
asymptotic state just described, the system evolves through
configurations where additional kinds of localized objects are
seen. The presence of the Nozaki-Bekki hole-pulse pair as initial
condition in the central part of Fig. 1 gives birth to a pair of
fronts which  replace the initial lateral plane-waves by new ones.
Interestingly, a different kind of localized object is seen to
form just where the initial hole-pulse pair was placed. A close-up
of it at $t=90$ is displayed in Fig. 2. It is a kind of coupled
maximum-maximum pair. The moduli of the two fields are superposed
in the central panel showing the full object. The lateral small
bumps are propagating waves that travel towards the central
maxima. Thus the center of the coherent structure acts as a wave
sink \cite{hohenbergsaarloos}.

\begin{figure}[h]
\begin{center}
\resizebox{80mm}{!}{\includegraphics{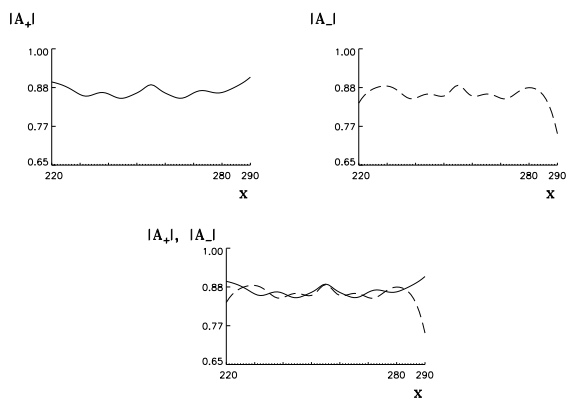}}
\end{center}
\caption[2]{ Snapshots
of part of the system in  Fig. 1 at $t= 90$ showing a
localized maximum--maximum wave sink.}
\label{figr2297c}
\end{figure}

In Figure 3 the spatiotemporal evolution of $|A_{+}(x,t)|$ and
$|A_{-}(x,t)|$ was obtained using as initial conditions a sharp
phase jump at the center of the system, with small random white
noise added. The parameter values are $ \alpha = 0.6 \,,\, \beta =
-1.4 \mbox{ and } \gamma = 0.7$. After a short time, the system
reaches a state dominated by branching hole--hole pair structures.
Lighter grey correspond to the maximum values of $|A_{\pm}(x,t)|$
and darker to the minima. The two big triangles correspond to
regions of constant modulus, that is, plane waves. The bottom
panels show $|A_{+}|$ and $|A_{-}|$ in a portion of the system at
these early times. Both are superposed in the central panel to
show the complete matching of the two fields.

At longer times, all the hole-hole pairs disappear from the
system, thus indicating that they are not stable objects at this
value of the parameters. The system decays to the same state as at
the end of Fig. 1: the dominant coherent structures
are the maximum-hole pairs.
\begin{figure}[h]
\begin{center}
\resizebox{80mm}{!}{\includegraphics{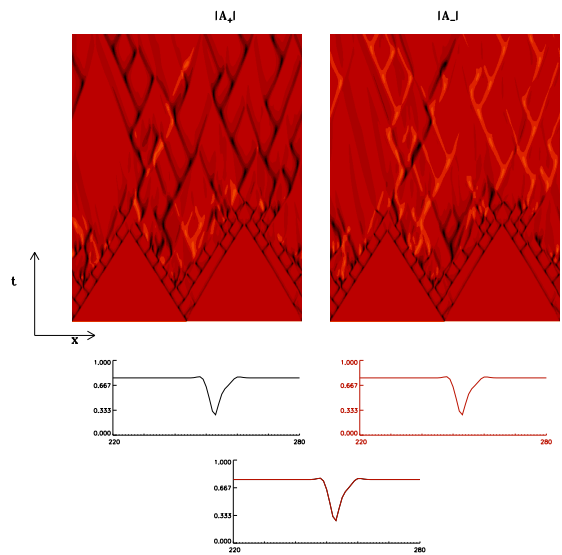}}
\end{center}
\caption[2]{ Spatiotemporal evolution of $|A_{+}(x,t)|$ and
$|A_{-}(x,t)|$ with
time running upwards from $0$ to $400$ and $x$ in the horizontal
direction from $0$ to $512$. Lighter red correspond to the maximum
value of $|A_{\pm}(x,t)|$ and darker to  the minimum. Bottom
panels (left and right) show one of the localized hole-hole
objects dominating the early dynamics. Central bottom panel
superpose them, showing its perfect matching. Both fields, then,
have exactly the same modulus around the core of the coherent
structure, as in the ansatz (\ref{transf}).}
\label{fig197c}
\end{figure}

\section{Exact solutions}

The different spatiotemporal evolutions shown in the previous
figures (1)--(3) are themselves interesting enough for a detailed
study. The localized objects appearing in the simulations are
clearly responsible for most of the complex dynamics in the
system. We can interpret some of the observed structures from a
simple ansatz:
 \begin{equation}
 A_+(x,t) = e^{i \varphi} A_-(x,t)
 \label{transf}
 \end{equation}
where $\varphi$ is constant, and $A_-(x,t)$ is any solution of the
single CGL equation:
 \begin{equation}
   \partial_t A_-  =  A_- + b \,  \partial_x^2 A_- - c \mid A_-\mid ^2
A_-  \>,
 \label{cglere}
 \end{equation}
 \noindent where $b = 1 +i\alpha$ and $c = (1+ \gamma) +
 i (1+ \gamma)\beta$ .

This simple ansatz gives us a rather rich set of exact solutions:
for each known analytical solution of the single CGL equation
(\ref{cglere}), there is a corresponding solution of the CCGL
equation set, in which  $A_-$ and $A_+$ have essentially the same
shape except for a constant global phase. In particular, hole,
pulse, shock, and front solutions are localized solutions
analytically known for the single equation
\cite{bekki83,bekki84,hohenbergsaarloos,conte93}, so that
hole-hole, pulse-pulse, shock-shock and front-front pairs are
immediately found as analytical solutions of the CCGL set. In
particular pulse-pulse and hole-hole structures are present in
Figs. 1 to 3, and turn out to be well described by the ansatz
(\ref{transf}).

It is worthwhile to note that the studies of \textsl{instability}
for these objects in the complex Ginzburg-Landau equation are
immediately translated into instability results for the paired
structures in CCGL equations.

\section{Conclusion}

In summary, we have shown numerically the existence of different
kinds of localized objects, responsible for the complex behavior
or solutions of the CCGL equations. Some of these objects can be
understood in terms of exact solutions arising from a simple
ansatz. A more detailed analysis is still needed, however. In
particular, the hole-maximum structure, which appears as the
dominant coherent structure at long times, can not be described by
our ansatz. In addition, much more work is needed in order to
establish the stability properties of the different objects, and
the nature of their interactions. In a recent
work\cite{conte99a,conte99b} new exact solutions of equation
\ref{theEqu} were obtained by using the Painlev\'{e} expansion
method. The authors describe these solutions as analogues of the
Nozaki-Bekki solutions \cite{bekki83,bekki84}. Comparison of these
solutions, different from the ansatz (\ref{transf}), with our
numerical results is under progress.

\vskip 0.2 cm
Financial support from DGICYT Projects PB94-1167 and
PB97-141-C02-01 is acknowledged. R.M. Acknowledge financial
support from CONICYT-Fondo Clemente Estable (Uruguay)


\end{document}